\documentclass[aps,pre,twocolumn,superscriptaddress,showpacs]{revtex4}
\usepackage{amssymb}
\usepackage{epsfig}
\usepackage{amsmath}
\usepackage{times}
\usepackage{multirow}
\begin{document}
\title{Percolation of partially interdependent networks under targeted attack}

\author{Gao-Gao Dong}\email{gago999@126.com}
\affiliation{Nonlinear Scientific for Research Center, Faculty of
Science, Jiangsu University, Zhenjiang, 212013, China}

\author{Jian-Xi Gao}
\affiliation{Department of Automation, Shanghai Jiao Tong
University, Shanghai, 200240, China}

\affiliation{Center for Polymer Studies and Department of Physics,
Boston University, Boston, MA 02215, USA}

\author{Li-Xin Tian}
\affiliation{Nonlinear Scientific for Research Center, Faculty of
Science, Jiangsu University, Zhenjiang, 212013, China}

\author{Rui-Jin Du}\email{dudo999@126.com}
\affiliation{Nonlinear Scientific for Research Center, Faculty of
Science, Jiangsu University, Zhenjiang, 212013, China}

\affiliation{College of Mathematics Science, Chongqing Normal
University, Chongqing, 401331, China}

\author{Ying-Huan He}
\affiliation{Nonlinear Scientific for Research Center, Faculty of
Science, Jiangsu University, Zhenjiang, 212013, China}

\begin{abstract}
The study of interdependent networks, and in particular the
robustness on networks, has attracted considerable attention. Recent
studies mainly assume that the dependence is fully interdependent.
However, targeted attack for partially interdependent networks
simultaneously has the characteristics of generality in real world.
In this letter, the comprehensive percolation of generalized
framework of partially interdependent networks under targeted attack
is analyzed. As $\alpha=0$ and $\alpha=1$, the percolation law is
presented. Especially, when $a=b=k$, $p_{1}=p_{2}=p$,
$q_{A}=q_{B}=q$, the first and second lines of phase transition
coincide with each other. The corresponding phase transition diagram
and the critical line between the first and the second phase
transition are found. We find that the tendency of critical line is
monotone decreasing with parameter $p_{1}$. However, for different
$\alpha$, the tendency of critical line is monotone increasing with
$\alpha$. In a larger sense, our findings have potential application
for designing networks with strong robustness and can regulate the
robustness of some current networks.
\end{abstract}


\maketitle
\section{Introduction}
Complex networks have been shown to exist in many different areas in
the real world and have been intensively studied in recent years.
However, almost all network research have been focused on properties
of a single network component that does not interact and depend on
other networks \cite{Watts1998, Bar1999, Albert2002, Cohen2000,
Callaway2000, Dorogovtsev2003, Song2005, Satorras2006,
Caldarelli2007, Havlin2010, Newman2010, Chen2011, Gao2010}. Such
situations rarely, if ever, occur in reality \cite{Gao2011,
Buldyrev2010, Parshani2010, Huang2011}. In 2010, Buldyrev et al.
\cite{Buldyrev2010} developed a theoretical framework for studying
the process of cascading failures in one-to-one correspondent
interdependent networks caused by random initial failure of nodes.
Surprisingly, they found that a broader degree distribution
increased the vulnerability of interdependent networks to random
failure, which is in contrast to the behavior of a single network.
Partially interdependent networks were investigated by Parshani et
al. \cite{Parshani2010}, they presented a theoretical framework for
studying this case in the same year. Their findings highlighted that
reducing the coupling strength could lead to a change from a first
to second order percolation transition. In 2011, Huang et al.
\cite{Huang2011} mainly demonstrated the robustness of fully
interdependent networks under targeted attack. The result implied
that interdependent networks were difficult to defend. However, when
real scenarios are considered, simultaneous attack on partially
interdependent networks intentionally is more general.

Motivated by the above, we develop a generalized framework to
comprehensively study the percolation of partially interdependent
networks that suffer targeted attack simultaneously. Likewise,
percolation law and condition of the first and the second phase
transitions have been analyzed for partially interdependent
networks. Furthermore, by applying generalized framework and
analyzing the percolation, as $\alpha$ of $W_{\alpha}(k_{i})$
varying, the corresponding percolation phase transition and critical
line between the first and the second phase transition are firstly
presented. Meanwhile, the robustness of two interdependent networks
can be comprehensively studied by using the generalized framework.

\section{The Model}
This model consists of two networks $A,B$ with the number of nodes
$N_{A},N_{B}$, and within each network, the nodes are connected with
degree distributions $P_{A}(k)$ and $P_{B}(k)$, respectively.
Suppose that the average degree of the network $A$ is $a$ and the
average degree of the network $B$ is $b$. In addition, a fraction
$q_{A}$ of network $A$ nodes depends on the nodes in network $B$ and
a fraction $q_{B}$ of network $B$ nodes depends on the nodes in
network $A$. That is, if node $A_{i}$ of network $A$ depends on node
$B_{j}$ of network $B$ and $B_{j}$ depends on node $A_{k}$ of
network $A$, then $k=i$. Consequently, when nodes from one network
fail, the interdependent nodes from the other network also fail.
This invokes an iterative cascade of failures in both networks. A
value $W_{\alpha}(k_{i})$ is assigned to each node, which presents
the probability that a node $i$ with $k_{i}$ links becomes inactive
by targeted-attack. We focus on the family of functions
\cite{Gallos2005}:
\begin{equation}
W_{\alpha}(k_{i})=\frac{k_{i}^{\alpha}}{\sum_{i=1}^{N}k_{i}^{\alpha}},
-\infty<\alpha<+\infty.
\end{equation}
When $\alpha\neq0$, nodes are attacked intentionally, while for
$\alpha=0$, nodes are removed in random.

We begin by studying the situation where both networks $A$ and $B$
are attacked simultaneously with probability $W_{\alpha}(k_{i})$
(Eq. (1)). Initially, $1-p_{1}$ and $1-p_{2}$ fraction of nodes are
intentionally removed from network $A$ and network $B$ respectively.
$p_{A}$ and $p_{B}$ are defined as the fraction of nodes belonging
to the giant components of networks $A$ and $B$. The remaining
fraction of network $A$ nodes after an initial removal of $1-p_{1}$
is $\psi_{1}'=p_{1}$, and the remaining fraction of network $B$
nodes after an initial removal of $1-p_{2}$ is $\phi_{0}'=p_{2}$.
The remaining functional part of network $A$ contains a fraction
$\psi_{1}=\psi_{1}'p_{A}(\psi_{1}')$ of network nodes. Accordingly,
the remaining fraction of network $B$ is
$\phi_{1}'=p_{2}[1-q_{B}(1-p_{A}(\psi_{1}')p_{1})]$, and the
fraction of nodes in the giant component of network $B$ is
$\phi_{1}=\phi_{1}'p_{B}(\phi_{1}')$. Then the sequence, $\psi_{n}$
and $\phi_{n}$, of giant components, and the sequence $\psi_{n}'$
and $\phi_{n}'$, of the remaining fraction of nodes at each stage of
the cascading failures, are constructed as follows:
\begin{equation}
\begin{split}
\psi_{1}'&=p_{1}, \psi_{1}=\psi_{1}'p_{A}(\psi_{1}'),\\
\phi_{0}'&=p_{2},
\phi_{1}'=p_{2}[1-q_{B}(1-p_{A}(\psi_{1}')p_{1})],\phi_{1}=\phi_{1}'p_{B}(\phi_{1}'),\\
\psi_{2}'&=p_{1}[1-q_{A}(1-p_{B}(\phi_{1}')p_{2})],
\psi_{2}=\psi_{2}'p_{A}(\psi_{2}'),\\
\phi_{2}'&=p_{2}[1-q_{B}(1-p_{A}(\psi_{2}')p_{1})],
\phi_{2}=\phi_{2}'p_{B}(\phi_{2}'),\\ \cdots\\
\psi_{n}'&=p_{1}[1-q_{A}(1-p_{B}(\phi_{n-1}')p_{2})],
\psi_{n}=\psi_{n}'p_{A}(\psi_{n}'),\\
\phi_{n}'&=p_{2}[1-q_{B}(1-p_{A}(\psi_{n}')p_{1})],
\phi_{n}=\phi_{n}'p_{B}(\phi_{n}').
\end{split}
\end{equation}
To determine the state of system (2) at the end of the cascading
process we look at $\psi_{n}', \phi_{n}'$ at the limit of
$n\rightarrow\infty$. The limit must satisfy the equations
$\psi_{n}'=\psi_{n+1}', \phi_{n}'=\phi_{n+1}'$ since eventually the
clusters stop fragmenting and the fractions of randomly removed
nodes at step $n$ and $n+1$ are equal. Denoting $\psi_{n}'=x$,
$\phi_{n}'=y$, we arrive at a system of two symmetric equations:
\begin{equation}
\begin{split}
x&=p_{1}[1-q_{A}(1-p_{B}(y)p_{2})],\\
y&=p_{2}[1-q_{B}(1-p_{A}(x)p_{1})].
\end{split}
\end{equation}
For equation (1), as $\alpha=0$, Fig. 1 show excellent agreement
between computer simulations of the cascade failures and the
numerical results obtained by solving system (2).
\begin{figure}
\centering \scalebox{0.6}[0.6]{\includegraphics{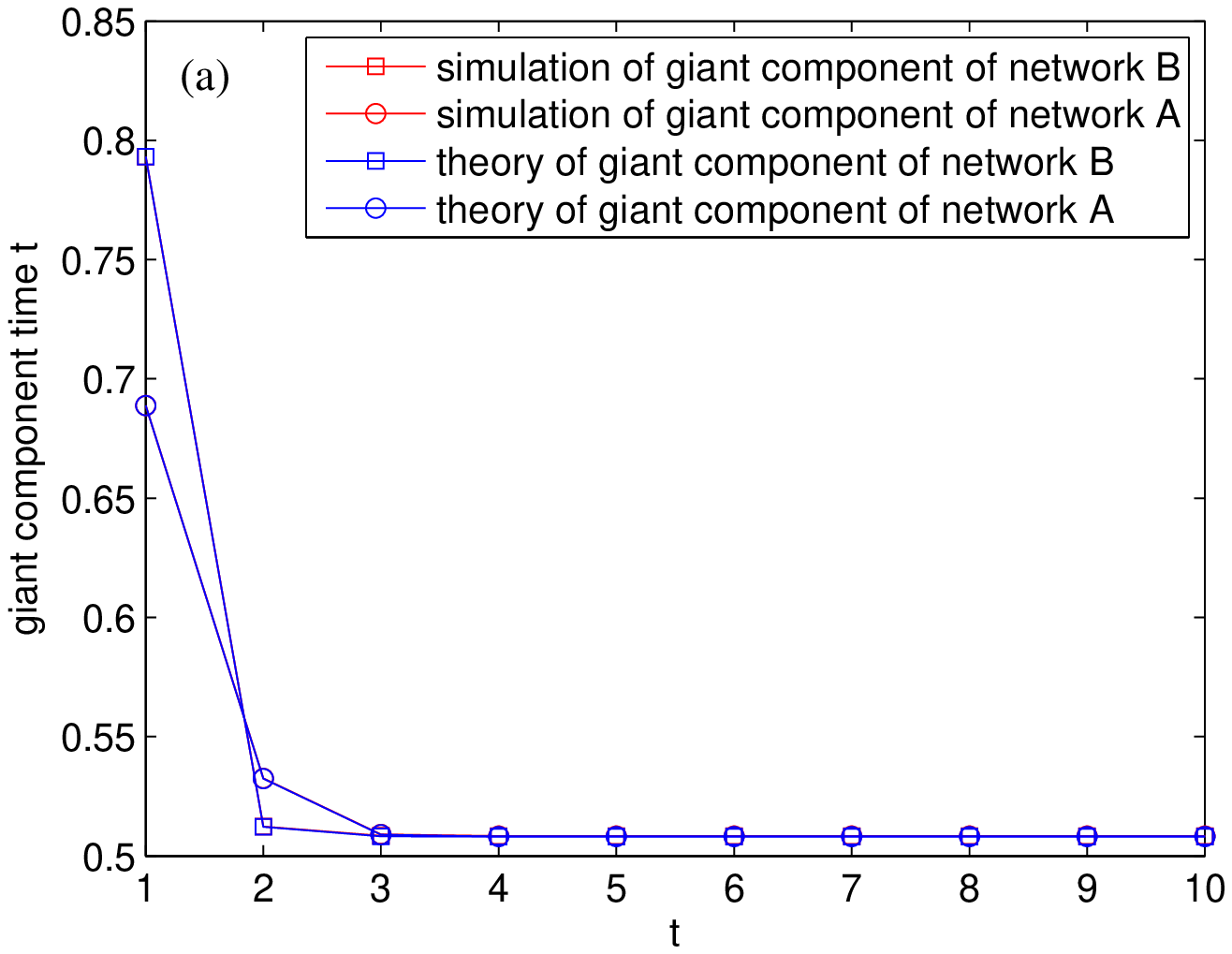}}
\scalebox{0.6}[0.6]{\includegraphics{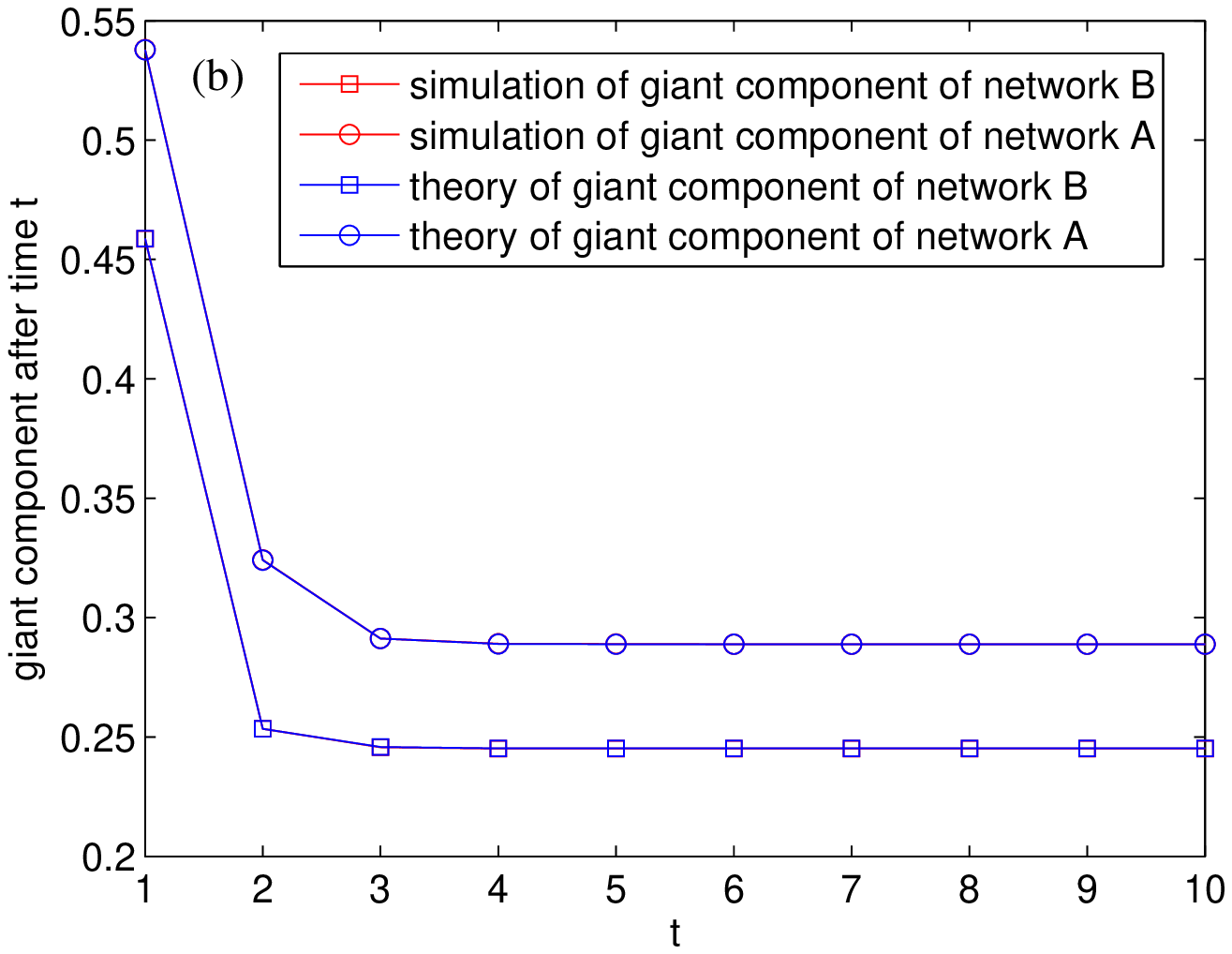}} \caption{(color
online) (a) Simulation results of the giant component of the two
fully interdependent ER networks after time $t$ cascading failures.
For each network $N=100,000$, $a=b=6$, $p_{1}=0.8$, $p_{2}=0.7$,
$q_{A}=q_{B}=1$. (b) Simulation results of the giant component of
the two partially interdependent ER networks after $t$ cascading
failures. For each network $N=100,000$, $a=b=6$, $q_{A}=0.6$,
$q_{B}=0.65$, $p_{1}=0.49$, $p_{2}=0.56$. All estimates are the
results of averaging over 10 realizations. $\psi_{\infty}$ and
$\phi_{\infty}$, the fraction of nodes in the giant components of
networks $A$ and $B$ separately, after a cascade of failures, the
simulations results for fully and partially interdependent networks
are identical with the theoretical values.}
\end{figure}
\section{Analytical Solution}
Here we discuss the exact analytical results when $\alpha=0$ and
$\alpha=1$. As in refs. \cite{Buldyrev2010, Parshani2010,
Newman2001, Newman2002}, we introduce the generating function of the
degree distribution $G_{A0}(\xi)=\Sigma_{k}P_{A}(k)\xi^{k}$, and the
generating function of the associated branching process,
$G_{A1}(\xi)=G_{A0}'(\xi)/G_{A0}'(1)$. The fraction of nodes that
belongs to the giant component after the removal of $1-p_{1}$ nodes
is \cite{Shao2008}:
\begin{equation}
p_{A}(p_{1})=1-G_{A0}[1-p_{1}(1-f_{A})],
\end{equation}
where $f_{A}=f_{A}(p_{1})$ satisfies a transcendental equation
\begin{equation}
f_{A}=G_{A1}[1-p_{1}(1-f_{A})].
\end{equation}

When $\alpha=0$, $W_{0}=\frac{1}{N}$, represents the random removal
of nodes. For the case of two Erd\"{o}s-R\'{e}nyi (ER)
\cite{Erd1959, Erd1960, Bol1985} networks with average degrees $a$
and $b$, we can easily get $p_{A}(x)=1-f_{A}$, $p_{B}(y)=1-f_{B}$,
and system (3) becomes
\begin{equation}
\begin{split}
x&=p_{1}[1-q_{A}+p_{2}q_{A}(1-f_{B})],\\
y&=p_{2}[1-q_{B}+p_{1}q_{B}(1-f_{A})].
\end{split}
\end{equation}
The fraction of nodes in the giant components of networks $A$ and
$B$, at the end of the cascading process are given by:
\begin{equation}
\begin{split}
\psi_{\infty}&=p_{1}(1-f_{A})[1-q_{A}+p_{2}q_{A}(1-f_{B})],\\
\phi_{\infty}&=p_{2}(1-f_{B})[1-q_{B}+p_{1}q_{B}(1-f_{A})].
\end{split}
\end{equation}
And $f_{A}$, $f_{B}$ can be expressed as:
\begin{equation}
\begin{split}
f_{A}(f_{B})&=\frac{1}{q_{B}}[\frac{1+q_{B}(p_{1}-1)}{p_{1}}-\frac{\ln{f_{B}}}{bp_{1}p_{2}(f_{B}-1)}],\\
f_{A}&\neq1; \forall f_{A}, f_{B}=1,\\
f_{B}(f_{A})&=\frac{1}{q_{A}}[\frac{1+q_{A}(p_{2}-1)}{p_{2}}-\frac{\ln{f_{A}}}{ap_{1}p_{2}(f_{A}-1)}],\\
f_{B}&\neq1; \forall f_{A}, f_{B}=1.
\end{split}
\end{equation}
In fact, for random attack ($\alpha=0$) on partially interdependent
networks simultaneously, the change between the first phase
transition and the second phase transition can be obtained by
regulating the coupling strength $q_{A}$ or $q_{B}$. Fig. 2 shows
that reducing the coupling strength lead to a change from the first
to second phase transition. And for different given $p_{1}$, the
critical line is also graphically founded ([Fig. 3]). From Fig. 3,
we can observe that for $\alpha=0$, the tendency of critical line is
monotone decreasing with increasing parameters $p_{1}$. And for
$p_{1}=0.6$, $p_{1}=0.7$, $p_{1}=0.8$ and $p_{1}=0.9$, the
corresponding percolation phase transitions are shown separately in
Figure 3.
\begin{figure}
\centering \scalebox{0.6}[0.6]{\includegraphics{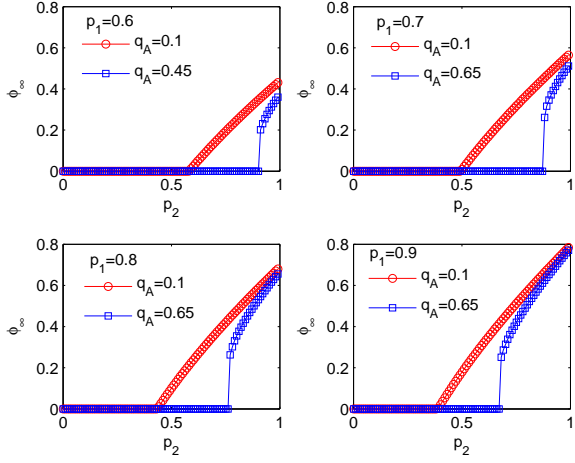}} \caption{
(color online) For given $p_{1}=0.6\thicksim0.9$, with the strong
coupling, the system undergoes the first phase transition at which
$\phi_{\infty}$, the fraction of nodes in the giant component of
network $B$, abruptly jumps from a finite value to zero. However,
with the weak coupling, the system undergoes the second phase
transition at which $\phi_{\infty}$, gradually approaches to zero.}
\end{figure}
\begin{figure}
\centering \scalebox{0.6}[0.6]{\includegraphics{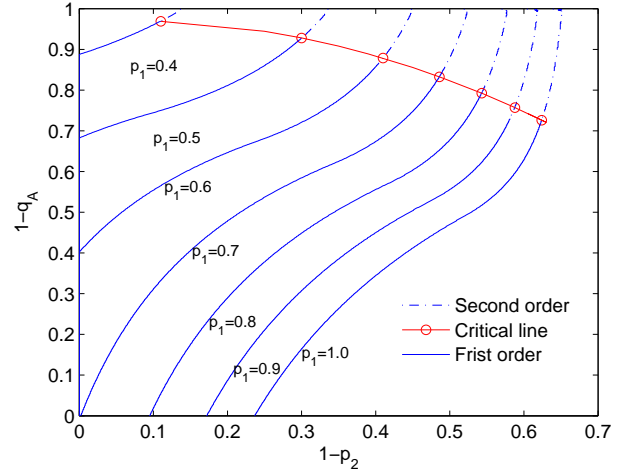}} \caption{
(color online) The percolation phase transition for network $B$ with
$a=b=3, q_{B}=0.7$, and $p_{1}=0.4\thicksim1.0$. The critical line,
as the boundary between the first and second phase transition, is
founded. The tendency of critical line is monotone decreasing with
$p_{1}$. Below the critical line, the system undergoes a first order
phase transition. Above the critical line, the system undergoes a
second order transition. }
\end{figure}

Especially when $a=b=k$, $p_{1}=p_{2}=p$, $q_{A}=q_{B}=q$, we arrive
to the equations:
\begin{equation}
f_{A}=f_{B}=f=e^{kp(f-1)[1-q+pq(1-f)]},  0\leq{f}<1,
\end{equation}
and
\begin{equation}
\psi_{\infty}=\phi_{\infty}=p(1-e^{-k\phi_{\infty}})[1-q+pq(1-e^{-k\phi_{\infty}})].
\end{equation}
Eq. (9) can be solved graphically as the intersection of a straight
line $y=f$ and a curve $y=e^{kp(f-1)[1-q+pq(1-f)]}$. When $p$ is
small enough the curve increases very slowly and does not intersect
with the straight line except at $f=1$ which corresponds to the
trivial solution. The condition for the first order transition
$(p=p^{I})$ is that the derivatives of the equations of system (9)
with respect to $f$:
\begin{equation}
1=f[kp^{I}(1-q)+2k(p^{I})^{2}q(1-f)].
\end{equation}
And solving system (9) for $f\rightarrow1$ yields the condition for
the second order transition $(p=p^{II})$:
\begin{equation}
kp^{II}(1-q)=1.
\end{equation}
The analysis of Eqs. (11) and (12) show that the first and second
order transition lines coincide each other:
\begin{equation}
p=p^{I}=p^{II}=\frac{1}{k(1-q)}.
\end{equation}
The critical values of $p_{c}$, $q_{c}$ for which the phase
transition changes from the first order to second order are obtained
when Eqs. (9), (11) and (12) are satisfied simultaneously, we get
the critical values are only related with average degree as
following:
\begin{equation}
\begin{split}
p_{c}&=\frac{k+1-\sqrt{2k+1}}{k},\\
q_{c}&=\frac{\sqrt{2k+1}+1}{2k}.
\end{split}
\end{equation}
The phase transition and critical line are graphically showed
respectively in Fig. 4. From Fig. 4, the tendency of critical line
descend with average degree $k$.
\begin{figure}
\centering \scalebox{0.6}[0.6]{\includegraphics{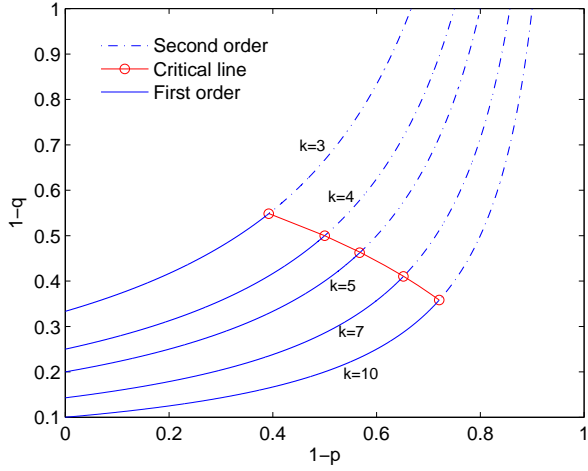}} \caption{
(color online) The phase transition is showed for $k=3, 4, 5, 7,
10$. The critical line is also graphically presented by Eq. (14).
Its tendency is monotone decreasing with increasing average $k$.}
\end{figure}

When $\alpha\neq0$, both of the two partially interdependent
networks are initially attacked with probability $W_{\alpha}(k_{i})$
(Eq. (1)). Initially, $1-p_{1}$ and $1-p_{2}$ fraction of nodes are
intentionally removed from network $A$ and network $B$ respectively.
The generating functions will be defined for network $A$ while
similar equations describe network $B$. According to Eq. (1),
$1-p_{1}$ fraction of nodes are removed from network $A$ but before
the links of the remaining nodes which connect to the removed nodes
are removed. The generating function of the nodes left in network
$A$ before removing the links to the removed nodes is
\cite{Huang2011, Newman2002, Shao2009}
\begin{equation}
G_{Ab}(x)=\sum_{k_{1}}P_{p_{1}}(k_{1})x^{k_{1}}=\frac{1}{p_{1}}\sum_{k_{1}}P(k_{1})h_{1}^{k_{1}^{\alpha}}x^{k_{1}},
\end{equation}
where the new degree distribution of the remaining fraction $p_{1}$
of nodes
$P_{p_{1}}(k_{1})=\frac{1}{p_{1}}P(k_{1})h_{1}^{k_{1}^{\alpha}}$,
and $G_{\alpha}(x)\equiv\sum_{k_{1}}P(k_{1})x^{k_{1}^{\alpha}}$,
$h_{1}\equiv{G_{\alpha}^{-1}(p_{1})}$. The generating function of
the new distribution of nodes left in network $A$ after the links to
the removed nodes are also removed is
\begin{equation}
G_{Ac}(x)=G_{Ab}(1-\tilde{p}_{1}+\tilde{p}_{1}x),
\end{equation}
where
$\tilde{p}_{1}=\frac{p_{1}N_{A}\langle{k_{1}(p_{1})}\rangle}{N_{A}\langle{k_{1}}\rangle}=\frac{\sum_{k_{1}}P(k_{1})k_{1}h_{1}^{k_{1}^{\alpha}}}{\Sigma_{k_{1}}P(k_{1})k_{1}}$
is the fraction of the original links that connect to the nodes
left, $\langle{k_{1}}\rangle$ is the average degree of the original
network $A$, $\langle{k_{1}(p_{1})}\rangle$ is the average degree of
remaining nodes before the links that are disconnected are removed.
Then we can find equivalent networks $A'$ and $B'$ with generating
functions $\tilde{G}_{A0}(x)$ and $\tilde{G}_{B0}(x)$, such that
after simultaneous random attack with $1-p_{1}$ and $1-p_{2}$
fractions of nodes, the new generating functions of nodes left in
$A'$ and $B'$ are the same as $G_{Ac}(x)$ and $G_{Bc}(x)$. That is,
the simultaneous targeted-attack problem on interdependent networks
$A$ and $B$ can be solved as simultaneous random-attack problem on
interdependent networks $A'$ and $B'$. By solving the equations
$\tilde{G}_{A0}(1-p_{1}+p_{1}x)=G_{Ac}(x)$,
$\tilde{G}_{B0}(1-p_{2}+p_{2}x)=G_{Bc}(x)$ and from Eq. (16) , we
can get
\begin{equation}
\begin{split}
\tilde{G}_{A0}(x)&=G_{Ab}(1-\frac{\tilde{p}_{1}}{p_{1}}+\frac{\tilde{p}_{1}}{p_{1}}x),\\
\tilde{G}_{B0}(x)&=G_{Bb}(1-\frac{\tilde{p}_{2}}{p_{2}}+\frac{\tilde{p}_{2}}{p_{2}}x).
\end{split}
\end{equation}
Simplified forms for $G_{Ab}(x)$, $G_{Ac}(x)$ and
$\tilde{G}_{A0}(x)$ from Eqs. (15), (16) and (17) exist when
$\alpha=1$,
\begin{equation}
G_{Ab}(x)=\frac{1}{p_{1}}\sum_{k_{1}}P(k_{1})h_{1}^{k_{1}}x^{k_{1}}=\frac{1}{p_{1}}G_{A0}(h_{1}x),
\end{equation}
\begin{equation}
G_{Ac}(x)=\frac{1}{p_{1}}G_{A0}(h_{1}(1-\tilde{p}_{1}+\tilde{p}_{1}x)),
\end{equation}
\begin{equation}
\tilde{G}_{A0}(x)=\frac{1}{p_{1}}G_{A0}(h_{1}(1-\frac{\tilde{p}_{1}}{p_{1}}+\frac{\tilde{p}_{1}}{p_{1}}x)),
\end{equation}
where $G_{A0}(x)$ is the original generating function of the network
$A$, $h_{1}=G_{A0}^{-1}(p_{1})$,
$\tilde{p}_{1}=\frac{G_{A0}'(h_{1})}{G_{A0}'(1)}h_{1}$. For ER
\cite{Erd1959, Erd1960, Bol1985} networks, we can also get
$p_{A}(x)=1-f_{A}$, $p_{B}(y)=1-f_{B}$, where
$p_{A}(x)=1-\tilde{G}_{A0}[1-x(1-f_{A})]$,
$f_{A}=\tilde{G}_{A1}[1-x(1-f_{A})]$, and system (3) becomes
\begin{equation}
\begin{split}
x&=p_{1}[1-q_{A}+p_{2}q_{A}(1-f_{B})],\\
y&=p_{2}[1-q_{B}+p_{1}q_{B}(1-f_{A})].
\end{split}
\end{equation}
\begin{figure}
\centering \scalebox{0.6}[0.6]{\includegraphics{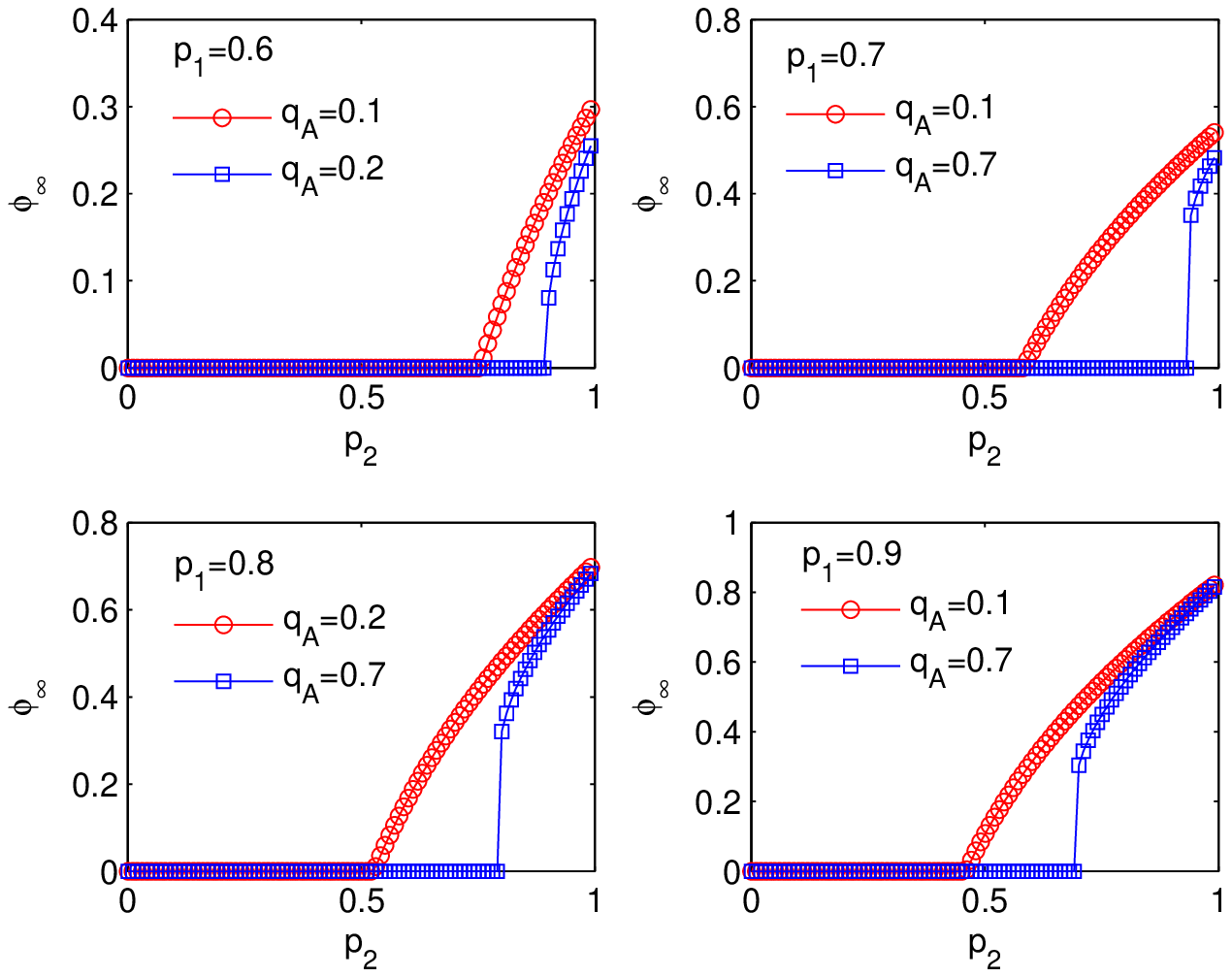}} \caption{
(color online) For given $p_{1}=0.6\thicksim0.9$,  the system
undergoes the first phase transition and the second phase transition
for network $B$ with strong and weak coupling.}
\end{figure}
The fraction of nodes in the giant components of networks $A$ and
$B$, respectively, at the end of the cascade process are then given
by:
\begin{equation}
\begin{split}
\psi_{\infty}=&p_{1}(1-f_{A})[1-q_{A}+p_{2}q_{A}(1-f_{B})],\\
\phi_{\infty}=&p_{2}(1-f_{B})[1-q_{B}+p_{1}q_{B}(1-f_{A})].
\end{split}
\end{equation}
And $f_{A}$, $f_{B}$ can be expressed as:
\begin{equation}
\begin{split}
f_{A}(f_{B})&=\frac{1}{q_{B}}[\frac{1+q_{B}(p_{1}-1)}{p_{1}}-\frac{\ln{f_{B}}}{bp_{1}p_{2}h_{2}^{2}(f_{B}-1)}],\\
f_{A}&\neq1; \forall{f_{A},f_{B}=1},\\
f_{B}(f_{A})&=\frac{1}{q_{A}}[\frac{1+q_{A}(p_{2}-1)}{p_{2}}-\frac{\ln{f_{A}}}{ap_{1}p_{2}h_{1}^{2}(f_{A}-1)}],\\
f_{B}&\neq1; \forall{f_{A},f_{B}=1}.
\end{split}
\end{equation}
Likewise, for targeted attack ($\alpha=1$) on partially
interdependent networks simultaneously, the phase transition changes
from a first order to a second order percolation transition, as the
coupling strength $q_{A}$ is reduced (Fig. 5). From Fig. 6, the
phase transition lines are graphically presented, and the critical
line descends with increasing parameters $p_{1}$.
\begin{figure}
\centering \scalebox{0.6}[0.6]{\includegraphics{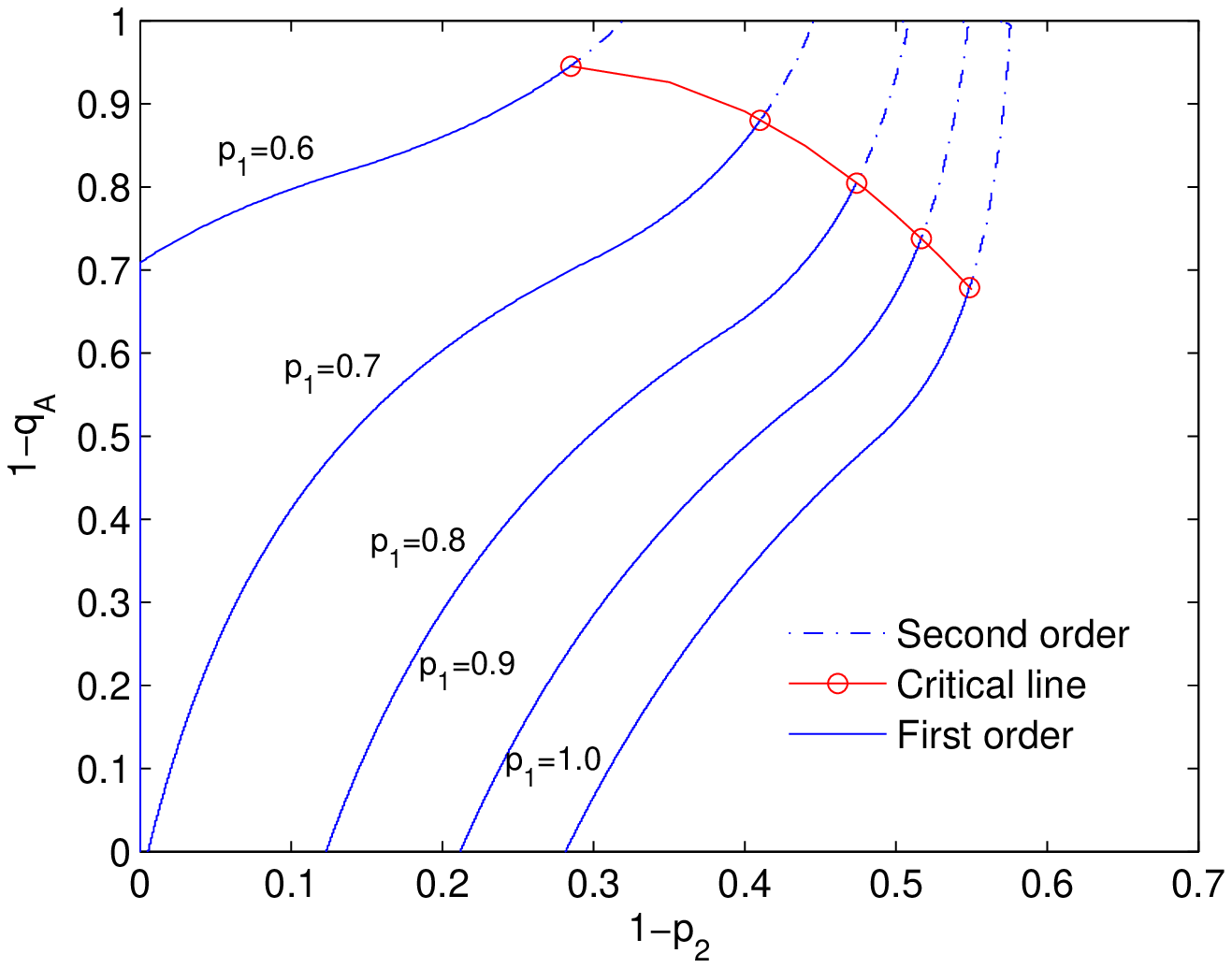}} \caption{
(color online) The percolation phase transition for network $B$ with
$a=3$, $b=4$, $q_{B}=0.7$ and $p_{1}=0.6\thicksim1.0$. The
corresponding phase transition lines are showed. Also, the critical
line is graphically found, and its tendency is monotone decreasing
with $p_{1}$. Below the critical line, the system undergoes a first
order phase transition. As we approach the critical point,
$\phi_{\infty}$ tends to 0. Above the critical line, the system
undergoes a second order transition.}
\end{figure}

Especially as $a=b=k$, $p_{1}=p_{2}=p$, $q_{A}=q_{B}=q$, we obtain
the following equations:
\begin{equation}
f_{A}=f_{B}=f=e^{-kph^{2}(1-f)[1-q+pq(1-f)]},
\end{equation}
and
\begin{equation}
\psi_{\infty}=\phi_{\infty}=p(1-e^{-kh^{2}\phi_{\infty}})[1-q+pq(1-e^{-kh^{2}\phi_{\infty}})],
\end{equation}
where $h=\frac{\ln{p}}{k}+1$. The condition for the first order
transition $(p=p^{I})$ is that the derivatives of system (24) with
respect to $f$:
\begin{equation}
1=f[kp^{I}h^{2}(1-q)+2k(p^{I})^{2}qh^{2}(1-f)],  0\leq{f}<1.
\end{equation}
And solving system (24) for $f\rightarrow1$ yields the condition for
the second order transition $(p=p^{II})$:
\begin{equation}
kp^{II}(1-q)h^{2}=1.
\end{equation}
From Eqs. (26) and (27), we have also found that the first order
transition line coincides with the second order transition line
($p$=$p^{I}$=$p^{II}$), the exact condition of phase transition is
as follows:
\begin{equation}
p(\frac{\ln{p}}{k}+1)^{2}=\frac{1}{k(1-q)}.
\end{equation}
\begin{figure}
\centering \scalebox{0.6}[0.6]{\includegraphics{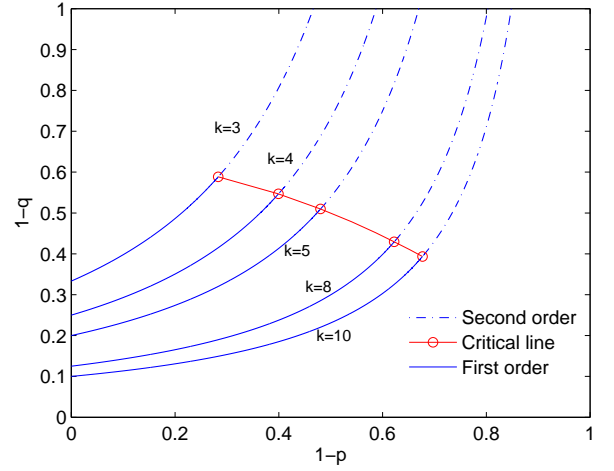}} \caption{
(color online) The corresponding phase transition is showed for
$k=3, 4 ,5, 8, 10$. The critical line is also graphically presented
from simulation. The tendency of critical line is found monotone
decreasing with average degree $k$}
\end{figure}
From Fig. 7, the simulation of phase transition is showed for Eq.
(28). Meanwhile, the critical line is also by simulation. From Fig.
7, the tendency of critical line is monotone decreasing with average
degree $k$.
\begin{figure}
\centering \scalebox{0.6}[0.6]{\includegraphics{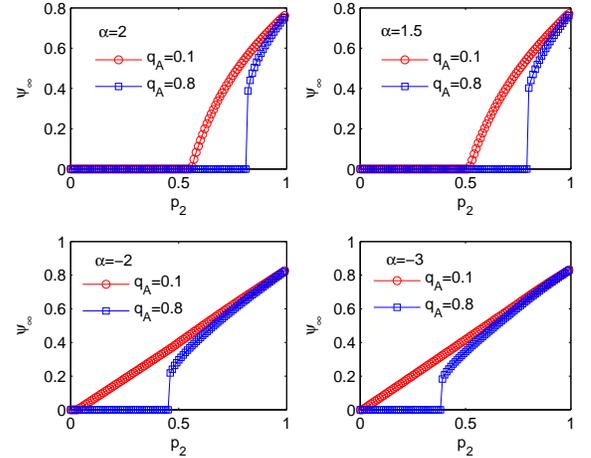}} \caption{
(color online) For given $\alpha=2, 1.5, -2, -3$,  the system
undergoes the first phase transition and the second phase transition
for network $A$ with strong and weak coupling.}
\end{figure}

Both networks A and B are attacked simultaneously with probability
$W_{\alpha}(k_{i})$, when $\alpha>0$, nodes with higher degree are
more vulnerable and those nodes are intentionally attacked, when
$\alpha<0$ node with higher degree have lower probability to fail.
For different values of $\alpha$, Fig. 8 reflect the relationship
between the phase transition and the coupling strength. Fig. 9 show
the phase transition lines for different given $\alpha>0$
respectively. The critical line is also founded and its tendency is
monotone increasing with $\alpha$.
\begin{figure}
\centering \scalebox{0.6}[0.6]{\includegraphics{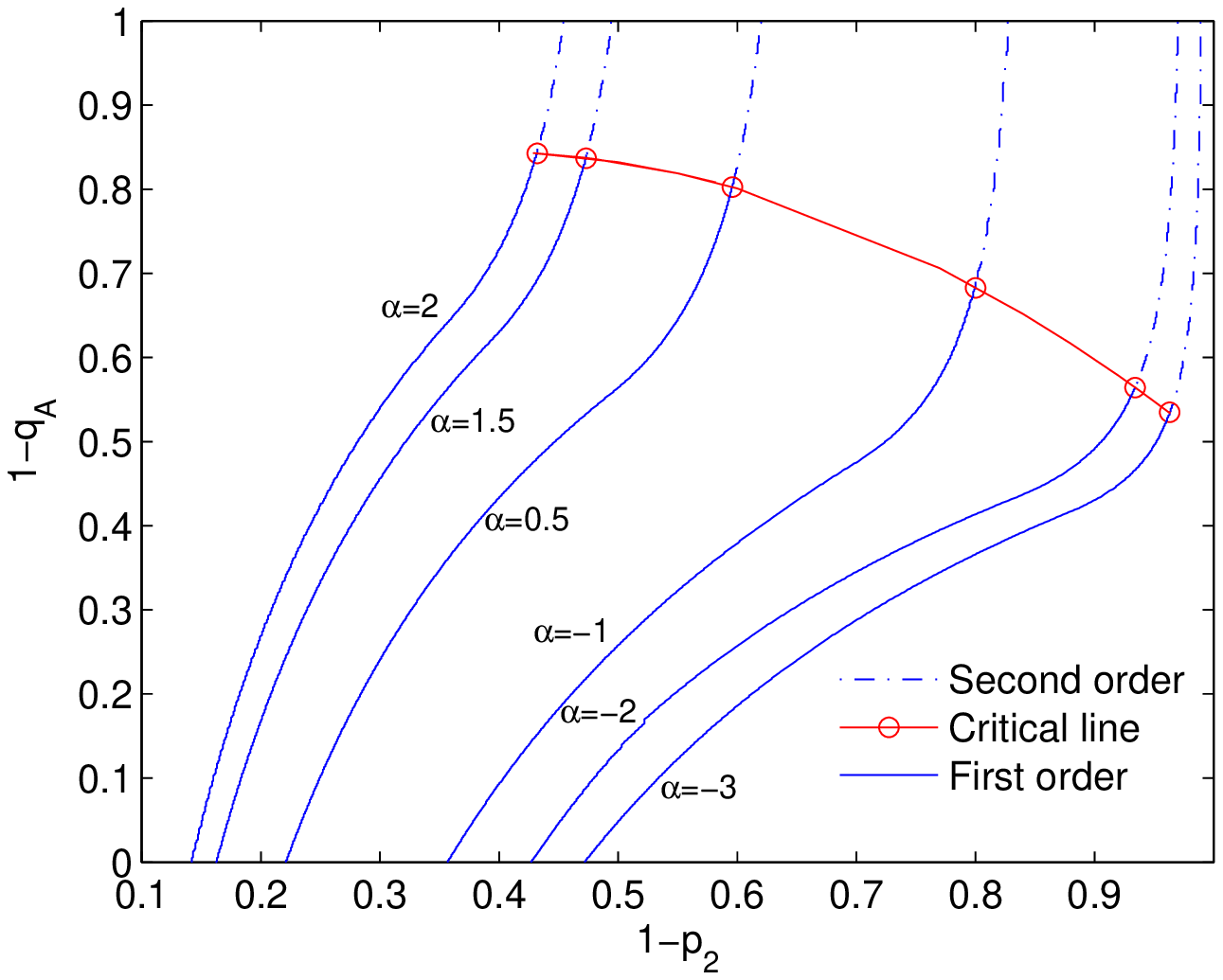}} \caption{
(color online) The percolation phase transition for network $A$ with
$a=3$, $b=4$, $q_{B}=0.7$, $p_{1}=0.8$ and $\alpha=-3, -2, -1, 0.5,
1.5, 2$. The corresponding phase transitions lines are showed. Also,
the critical line is graphically found, and its tendency is monotone
increasing with $\alpha$. Below the critical line, the system
undergoes a first order phase transition. As we approach the
critical point, $\psi_{\infty}$ tends to 0. Above the critical line,
the system undergoes a second order transition.}
\end{figure}

\section{Conclusions}
For interdependent networks in real scenario, two factors are
necessary to be considered: partial coupling and targeted attack. A
general framework is proposed to investigate the percolation of
partially interdependent networks that suffer targeted-attack
simultaneously. The percolation of partially interdependent networks
under targeted attack is comprehensively analyzed. As $\alpha=0$ and
$\alpha=1$, the percolation law is described detailedly .
Especially, for $a=b=k$, $p_{1}=p_{2}=p$, $q_{A}=q_{B}=q$, the first
and second lines of phase transition coincide with each other. And,
the tendency of critical line monotone decreasing with average
degree $k$. we show both analytically and numerically that reducing
the coupling between the networks leads to a change from a first to
a second order phase transition at a critical line. The tendency of
critical line is monotone decreasing with parameter $p_{1}$.
However, for different $\alpha$, the percolation phase transition is
also graphically demonstrated and critical line is monotone
increasing with $\alpha$. Therefore, our finding should be worth
considering in designing robust network considering.

\begin{acknowledgments}
This work is funded by the National Natural Science Foundation of
China (Grant Nos. 91010011, 71073072, 51007032), the Natural Science
Foundation of Jiangsu Province (Grant No. 2007098), the National
Natural Science (Youth) Foundation of China (Grant No. 10801140),
the Graduate innovative Foundation of Jiangsu Province CX10B\_272Z
and the Youth Foundation of Chongqing
Normal University (Grant No. 10XLQ001).\\
\end{acknowledgments}

\end{document}